\documentclass[aps,prb,twocolumn,superscriptaddress,showpacs]{revtex4}
\usepackage{amsmath}
\usepackage{amssymb}
\usepackage{graphicx}
\usepackage{dcolumn}
\usepackage{bm}

\begin{document}

\title{Extremal transmission through a microwave photonic crystal and the observation of edge states in a rectangular Dirac billiard}

\author{S.~Bittner}
\affiliation{Institut f{\"u}r Kernphysik, Technische Universit{\"a}t
Darmstadt, D-64289 Darmstadt, Germany}

\author{B.~Dietz}
\email{dietz@ikp.tu-darmstadt.de}
\affiliation{Institut f{\"u}r Kernphysik, Technische Universit{\"a}t
Darmstadt, D-64289 Darmstadt, Germany}

\author{M.~Miski-Oglu}
\affiliation{Institut f{\"u}r Kernphysik, Technische Universit{\"a}t
Darmstadt, D-64289 Darmstadt, Germany}

\author{A.~Richter}
\email{richter@ikp.tu-darmstadt.de}
\affiliation{Institut f{\"u}r
Kernphysik, Technische Universit{\"a}t Darmstadt, D-64289 Darmstadt,
Germany}
\affiliation{$\rm ECT^*$, Villa Tambosi, I-38123 Villazzano (Trento), Italy}

\date{\today}

\begin{abstract}
This article presents experimental results on properties of waves propagating in an unbounded and a bounded photonic crystal consisting of metallic cylinders which are arranged in a triangular lattice. First, we present transmission measurements of plane waves traversing a photonic crystal. The experiments are performed in the vicinity of a Dirac point, i.e., an isolated conical singularity of the photonic band structure. There, the transmission shows a pseudodiffusive $1/L$ dependence, with $L$ being the thickness of the crystal, a phenomenon also observed in graphene. Second, eigenmode intensity distributions measured in a microwave analog of a relativistic Dirac billiard, a rectangular microwave billiard that contains a photonic crystal, are discussed. Close to the Dirac point states have been detected which are localized at the straight edge of the photonic crystal corresponding to a zigzag edge in graphene. 
\end{abstract}

\maketitle
\section{\label{intr}Introduction}
The experimental realization of graphene~\cite{Novoselov2004}, a one-atom-thick sheet of carbon atoms arranged in a honeycomb crystal lattice~\cite{Wallace1947,Boehm1994}, attracted a lot of interest due to its peculiar electronic band structure which exhibits a linear dispersion at low energies. There the band structure closely resembles the spectrum of a relativistic massless spin one-half particle~\cite{Castro2009}. As a consequence, even though the Fermi velocity of electrons in graphene is only $c/300$ with $c$ the speed of light, the electronic excitations are governed by the Dirac equation. The peculiar electronic properties lead to interesting physical phenomena~\cite{Beenakker2008,Castro2009}. 

\par
\begin{figure}[!t]
\centering
{\includegraphics[height=5cm]{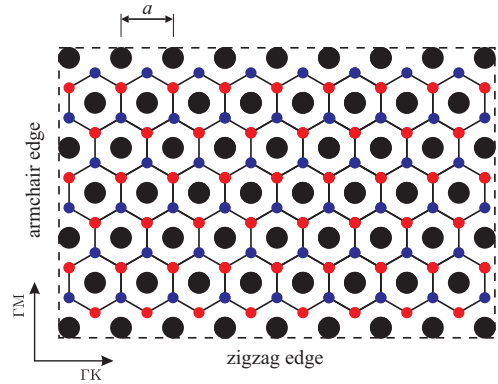}}
\caption{(Color online) Schematic view of the triangular lattice of metallic cylinders (black). The voids between the cylinders are marked by colored circles. For graphene the red (bright) and the blue (dark) voids correspond to the two atoms which generate the two independent triangular sublattices. The arrows indicate the two directions $\Gamma{\rm M}$ and $\Gamma{\rm K}$ considered in the transmission experiments. The radius of the cylinders equals $R=0.25\, a$, where $a$ is the lattice constant. The dashed lines indicate the positions of the metallic walls used in the experiments described in Sec.~\ref{Edge} to realize a microwave Dirac billiard.}
\label{fig1}
\end{figure}

\begin{figure}[!b]
{\includegraphics[width=\linewidth]{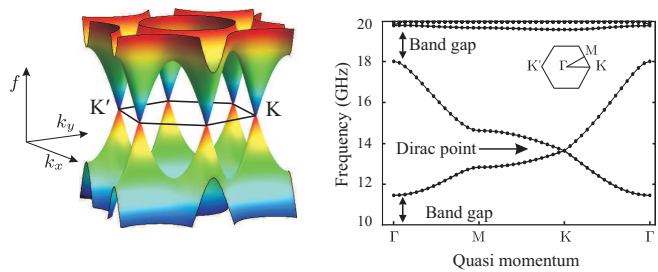}}
\caption{(Color online) Left: band structure $f(k_x,k_y)$ of the triangular lattice shown in Fig.~\ref{fig1} as function of the quasi-momentum components $(k_x,k_y)$. It was computed by numerically solving the scalar Helmholtz equation for the electric field strength imposing the Dirichlet boundary condition at the metallic cylinders. Shown are the first and the second band. Blue (darkest) color corresponds to the lowest, red (brightest) color to the largest values of $f$ within a band. The hexagonal Brillouin zone with the two non-equivalent $K$ points ${\bf K}$ and ${\bf K^\prime}$ is indicated. Right: band structure computed numerically by varying the quasi-momentum vector along the path $\Gamma$MK$\Gamma$ defining the boundary of the irreducible Brillouin zone depicted together with the Brillouin zone in the inset. Adopted from Ref.~\cite{Bittner2010}.
} 
\label{fig2}
\end{figure}

  
The linear dispersion relation is entirely due to the threefold rotational symmetry of the hexagonal graphene lattice~\cite{Slonczewski1958}. Therefore, graphene with its Dirac spectrum is not an exception. In general, systems comprising waves that propagate in a spatially periodic potential with the same symmetry properties can posses a Dirac spectrum. One example are so-called photonic crystals~\cite{Joannopoulos2008}, optical analogs of the common ones. It was shown theoretically in Ref.~\cite{Raghu2008} that under certain conditions photonic crystals with a triangular lattice geometry, depicted schematically in Fig.~\ref{fig1}, can exhibit the linear Dirac dispersion relation.
The dependence of the frequencies $f$ of propagating electromagnetic waves on the components $(k_x,k_y)$ of the quasi-momentum vector exhibits a band structure which is similar to that for the electronic energies in graphene. The left hand side of Fig.~\ref{fig2} shows the first two bands obtained for a two-dimensional photonic crystal consisting of metallic cylinders of radius $R=0.25 a$, 
where $a$ is the lattice constant. At the cylinders, the electric field strength vanishes, that is, obeys the Dirichlet boundary condition. The band structure was calculated by numerically solving the scalar Helmholtz equation for the electric field strength using the finite difference method described in Ref.~\cite{Smirnova2002}. The two bands touch each other at the corners of the hexagonal Brillouin zone depicted in Fig.~\ref{fig2} where they have the shape of cones. Thus, there the frequencies of propagating modes depend linearly on the distance of the quasi momentum from the touch points. The Brillouin zone comprises two non-equivalent corners ${\bf K}$ and ${\bf K^\prime}$, so-called {\it K} points, that cannot be mapped onto each other by basis vectors of the reciprocal lattice. The band diagram closely resembles that of the electronic energies in graphene~\cite{Katsnelson2007}. 
Accordingly, as in graphene~\cite{Novoselov2005} the touch points are referred to as Dirac points. 

The photonic crystal also exhibits a hexagonal configuration. In fact, three neighboring cylinders forming a triangular cell of the photonic crystal lattice shown in Fig.~\ref{fig1} host quasibound states~\cite{Gaspard1989} localized at their center. The cell thus can be considered as an open resonator and the triangular lattice can be regarded as being built of coupled resonators, the voids between the cylinders acting as the atoms forming the two independent triangular sublattices that generate the honeycomb structure. In Fig.~\ref{fig1} the voids corresponding to the two sublattices are marked with blue and red circles, respectively. The void structure terminates with a zigzag edge along the $\Gamma{\rm K}$ and with an armchair edge along the $\Gamma{\rm M}$ direction, respectively. These are the two most common types of boundaries arising from the hexagonal structure of graphene. 

In the vicinity of the Dirac points the Schr{\"o}dinger equation for the electrons in graphene~\cite{Ando2005,Brey2006,Castro2009} and the Helmholtz equation governing the electromagnetic wave propagation in the photonic crystal~\cite{Raghu2008} reduce to two copies of the Dirac equation for massless relativistic spin one-half particles, one copy holding around ${\bf K}$ and the other around ${\bf K^\prime}$. Thus, photonic crystals can be used to investigate properties of graphene, as was done in analog experiments with microwave photonic crystals~\cite{Zandbergen2010,Kuhl2010a,Bittner2010} and also with sonic crystals~\cite{Zhang2008}. There phenomena directly related to the linear dispersion relation around a Dirac point were investigated. We consider here two of them, the extremal transmission and the formation of edge states~\cite{Sepkhanov2007,Tworzydlo2006,Brey2006}. Both are properties of the waves propagating through a photonic crystal and crucially depend on the structure of the crystal edges. 

The Dirac dispersion relation affects the transport properties of photonic crystals in a remarkable way. It was shown theoretically in Refs.~\cite{Sepkhanov2007,Sepkhanov2008} that near the Dirac point the intensity of electromagnetic waves transmitted through a photonic crystal is inversely proportional to the thickness of the crystal. This is different from the ballistic behavior and the exponential decay observed, respectively, in a band region and a gap region, but rather reminiscent of the diffusion behavior of waves traveling through a disordered medium. A similar behavior was predicted for the conductivity of an ideal graphene strip~\cite{Tworzydlo2006,Katsnelson2006a}. We present below results on the experimental investigation of the transmission of electromagnetic waves traversing a microwave photonic crystal. The measurements (see Sec.~\ref{Extrem}) were performed close to a Dirac point, which was located as described in Ref.~\cite{Bittner2010}.

In further experiments (see Sec.~\ref{Edge}) we investigated edge effects~\cite{Nakada1996,Castro2009,Wurm2011} on the properties of the wave functions~\cite{Ponomarenko2008,Wurm2009} of a microwave analog of Dirac billiards. These experiments were motivated by the results obtained for the spectral properties of graphene flakes, also called graphene billiards~\cite{Wurm2011}, which have been shown to depend on the geometry of their edges~\cite{Ponomarenko2008,Wurm2009,Libisch2009}. For a graphene flake terminated with armchair edges the density of states vanishes at the Dirac point such that a gap opens up in the band structure, whereas zigzag edges yield a nonvanishing density. Indeed, in the vicinity of the Dirac point graphene flakes which terminate with zigzag edges or with both armchair and zigzag edges exhibit edge states, which are localized close to the zigzag parts of its boundary~\cite{Kobayashi2005,Kobayashi2006,Tao2011}. This property is attributed to the boundary conditions~\cite{Akhmerov2008,Wurm2011} which differ for both types of edges. To check the existence of such states in photonic crystals of finite size, one with the structure shown in Fig.~\ref{fig1} was enclosed by a metallic frame and wave functions were measured in the vicinity of a Dirac point. While in graphene flakes electrons are confined by a potential barrier, the walls of the metallic box prevent microwaves from leaving the photonic crystal and correspond to an infinitely high potential barrier in the quantum analog. 

\section{Experiments with the microwave photonic crystal\label{Model}}
Analog experiments with flat, cylindrical microwave cavities have been performed to model generic properties of quantum chaotic systems and of chaotic quantum scattering systems~\cite{Sridhar1991,Richter1999,StoeckmannBuch2000,Dietz2007b,Dietz2010,Kuhl2005a,RMP}. This is possible in a range $f\leq f_{\rm max}$ of the excitation frequency $f$, where the wave length is longer than twice the height of the cavity. There the electric field vector $\vec{E}$ is perpendicular to the bottom and the top plates of the cavity and obeys the Dirichlet boundary condition at the side walls of the cavity. Accordingly, the vectorial Helmholtz equation reduces to a scalar one which is mathematically equivalent to the two-dimensional stationary Schr{\"o}dinger equation for a non-relativistic quantum billiard of corresponding shape~\cite{Sridhar1991,Richter1999,StoeckmannBuch2000}. Due to this analogy flat microwave cavities are commonly called microwave billiards. We used microwave photonic crystals consisting of metallic cylinders which were arranged in a triangular lattice (see Fig.~\ref{fig1}) and squeezed between two metal plates to model single particle properties of graphene. At the cylinders the electric field strength vanishes, i.e., it obeys the Dirichlet boundary condition. Thus, for frequencies $f<f_{\rm max}$ the setup corresponds to a quantum multiple-scattering system. The experiments are described in Sec.~\ref{Extrem} and Sec.~\ref{Edge}, respectively.
\subsection{Extremal transmission at the Dirac frequency \label{Extrem}}
\begin{figure}[ht]
{\includegraphics[width=\linewidth]{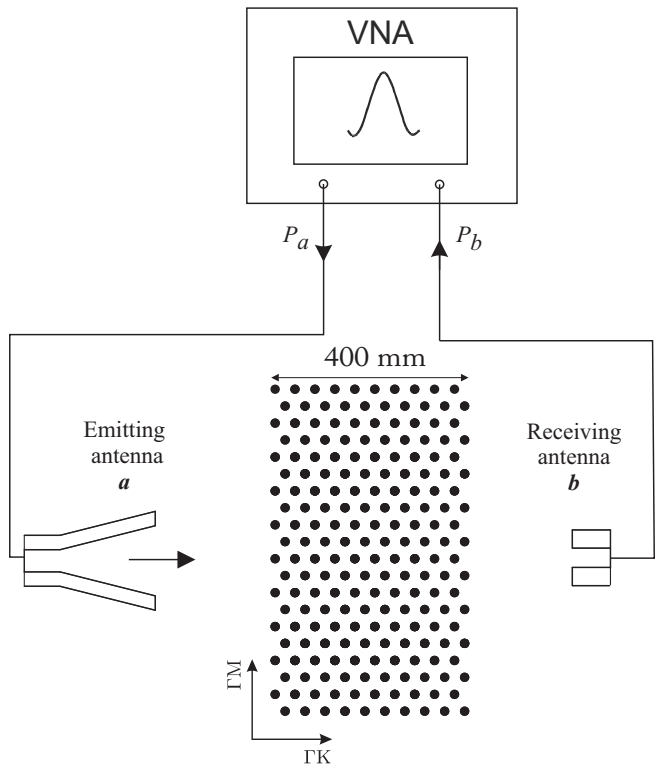}}
\caption{Experimental setup (not to scale). The photonic crystal consisted of $960$ copper cylinders arranged in a triangular lattice. A microwave horn antenna with an opening angle of $2\Delta\theta =28^\circ$ was used as emitting antenna and the signal transmitted through the crystal was received by a waveguide antenna. The crystal was placed symmetrically with respect to the antennas and oriented such that normal incidence of the emitted waves coincided with the $\Gamma {\rm K}$ direction, in distinction to the experiments described in Ref.~\cite{Bittner2010} where it corresponded to the $\Gamma {\rm M}$ direction.} 
\label{fig3}
\end{figure}

For the experimental investigation of the transmission properties of a photonic crystal with a Dirac spectrum we used the setup shown in Fig.~\ref{fig3}.
The photonic crystal consisted of 960 cylinders with radius $R=5$~mm and height $h=8$~mm. To ensure a good electrical contact and thus reproducibility of the measurement each cylinder was screwed to the top and the bottom plate. The resulting photonic crystal contained 40 rows of 24 cylinders and had side lengths of, respectively, $400$~mm and $859$~mm. The height of the cylinders defined the maximal frequency $f_{\rm max}=18.75$~GHz up to which the system was two dimensional. A two-dimensional horn antenna was used as emitting antenna and a waveguide antenna received waves transmitted through the crystal. Both were tightly screwed to the top plate and attached to waveguide-to-coaxial adapters. The transmission spectra were obtained with an Agilent PNA-L N5230A vectorial network analyzer (VNA). It measures the ratio of the power $P_b$ of the signal received at the waveguide antenna and the power $P_a$ of that emitted from the horn antenna and their relative phase. The modulus square of the scattering matrix element $S_{ba}$ from antenna $a$ to antenna $b$ is given as
\begin{equation}
 \lvert S_{ba}(f)\rvert^2=\frac{P_b}{P_a}.
\end{equation}
The horn antenna emits a beam with divergence $2\Delta\theta$, where $\Delta\theta=14^\circ$ with respect to the direction indicated by the arrow in Fig.~\ref{fig3}, which impinges the crystal situated in its near field. The orientation of the crystal was chosen such that normal incidence coincided with the $\Gamma{\rm K}$ direction (see Fig.~\ref{fig2}). In contrast, in the experiments described in Ref.~\cite{Bittner2010} it corresponded to the $\Gamma{\rm M}$ direction. 

\begin{figure}[!t]
{\includegraphics[width=\linewidth]{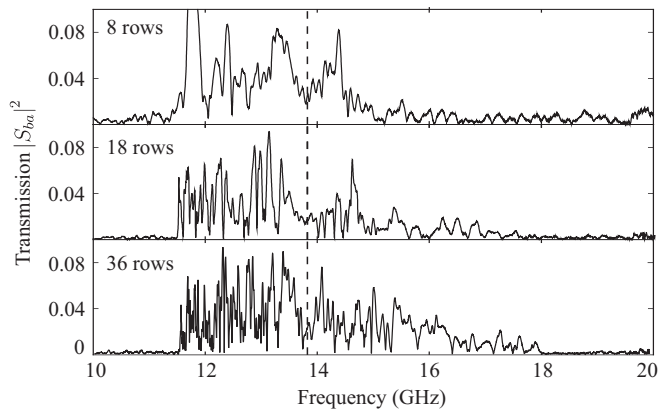}}
\caption{Examples of transmission spectra measured with the setup shown in Fig.~\ref{fig3} in $\Gamma {\rm K}$ direction through a lattice with, respectively, $8$, $18$, and $36$ rows of cylinders. The dashed line marks the Dirac frequency.} 
\label{fig4}
\end{figure}

The transmission spectra shown in Fig.~\ref{fig4} were taken for different configurations of the experimental setup.
First, the transmission spectrum was measured for the setup with 40 rows and then, step by step one row was removed from each armchair edge of the crystal.  The three panels show exemplarily the spectra measured for crystals consisting of $8$, $18$, and $36$ rows of cylinders, respectively. In all spectra the transmission is suppressed below $11.8$~GHz. There, the band structure exhibits a gap (see right panel of Fig.~\ref{fig2}), and electromagnetic waves cannot propagate in the photonic crystal. Furthermore, the transmission spectra show a characteristic dip around the Dirac frequency $f_D=13.797$~GHz, which was determined experimentally in Ref.~\cite{Bittner2010} and is marked in Fig.~\ref{fig4} with a dashed line. 
\begin{figure}[!ht]
\begin{center}
 {\includegraphics[width=0.85\linewidth]{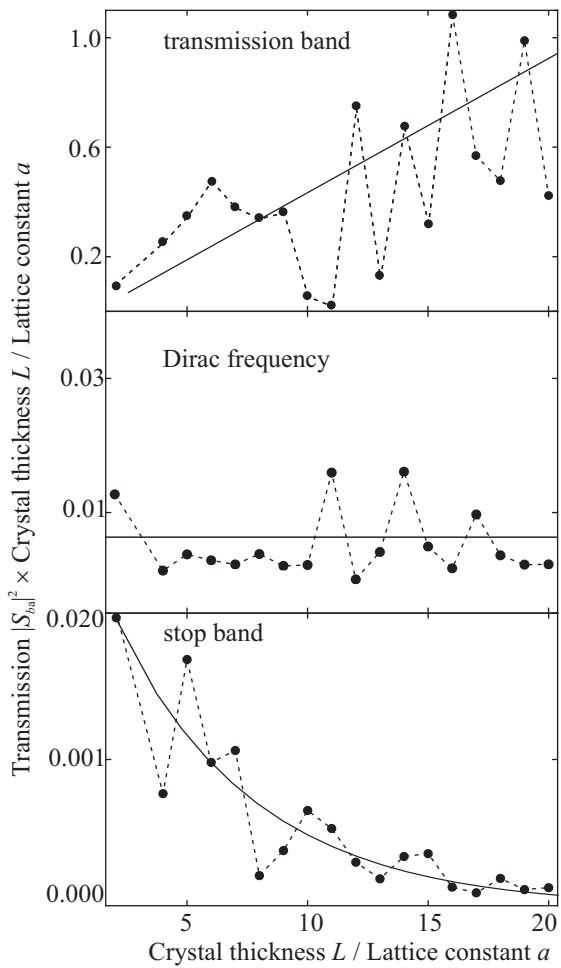}}
\end{center}
\caption{The product of the transmission $\vert S_{ba}(f)\vert^2$ through the photonic crystal and its thickness $L$ in units of the lattice constant $a$ as function of $L$ for different frequencies, from top to bottom for 13.18 GHz, 13.797 GHz, and 18.7 GHz. The filled circles are experimental data. They are joined with dashed lines to guide the eye. The solid lines result from the fits of a linear, a constant and an exponential function, respectively, to the data points.} 
\label{fig5}
\end{figure}
\begin{figure}[!ht]
\begin{center}
 {\includegraphics[width=0.85\linewidth]{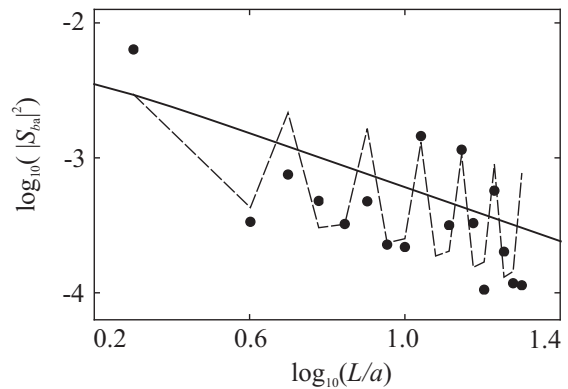}}
\end{center}
\caption{Double-logarithmic plot of the transmission $\vert S_{ba}(f)\vert^2$ through the photonic crystal as function of $L$ at the Dirac frequency $f_D=13.797$~GHz. The filled circles are experimental data. The solid line shows the prediction for the average behavior obtained by evaluating Eqs.~(4)-(6) of Ref.~\cite{Sepkhanov2008}. The fit of Eq.~(\ref{eq:1}) describing the oscillations~\cite{Zhang2008} yields the dashed line.} 
\label{fig5a}
\end{figure}

To reveal the $1/L$ scaling, with $L$ denoting the crystal thickness, of the transmitted power $\vert S_{ba}(f)\vert^2$ predicted near the Dirac point, we show in Fig.~\ref{fig5} the quantity $(L/a)\vert S_{ba}(f)\vert^2$ versus $L/a$ for three different frequencies. In all three panels the filled circles are the experimental data. They are joined by dashed lines to guide the eye. The solid curves depict the average behavior. 
The top panel shows $(L/a)\lvert S_{ba}\rvert^2$ versus $L/a$ for $f=13.18$~GHz, where the transmission is nonvanishing. In this frequency regime $L\lvert S_{ba}\rvert^2$ grows linearly with $L$, that is, the transmission $\lvert S_{ba}\rvert^2$ itself does not depend on $L$, as expected in band regions. The oscillations around the mean are attributed to standing waves arising between the antennas and the interfaces of the photonic crystal. The bottom panel of Fig.~\ref{fig5} shows the variation of $(L/a)\lvert S_{ba}\rvert^2$ with $L/a$ for $f=18.7$~GHz. This frequency lies within a stop band. As expected the transmission decreases exponentially with the crystal thickness $L$. 
The middle panel shows $(L/a)\lvert S_{ba}\rvert^2$ versus $L/a$ at the Dirac frequency $f_D=13.797$~GHz. The average behavior of $L\lvert S_{ba}\rvert^2$ is constant over a  wide range of the crystal thickness. Thus, the transmission of plane waves excited with the Dirac frequency and traveling in directions close to the $\Gamma {\rm K}$ direction, i.e., with $|\vec k_{||}|\approx 0$, where $\vec k_{||}$ is the wave vector component parallel to the interface of the crystal, scales with $1/L$. This is similar to the diffusion behavior of waves propagating through a random medium. As was pointed out in Ref.~\cite{Wang2009a} this phenomenon of extremal transmission is due to the linear Dirac dispersion relation. Indeed, in the vicinity of a Dirac point the local density of states vanishes only exactly at the Dirac frequency, whereas in a band gap it is zero for a finite range of excitation frequencies. As a consequence, the observability of the $1/L$ scaling crucially depends on the orientation of the photonic crystal with respect to the incident plane waves. When emitting plane waves excited with the Dirac frequency  in $\Gamma {\rm M}$ direction, as was done in previous experiments~\cite{Zandbergen2010,Bittner2010}, the band structure has a partial band gap (see right panel of Fig.~\ref{fig2}). Accordingly, the transmission $\lvert S_{ba}\rvert^2$ decreases exponentially with the crystal thickness $L$, that is, the decay length of the evanescent waves is finite. In contrast, it is infinite when exciting the Dirac point, e.g., by sending plane waves at the Dirac frequency in $\Gamma {\rm K}$ direction onto the crystal. 

The $1/L$ scaling near a Dirac point has been predicted for photonic crystals in Ref.~\cite{Sepkhanov2007}. For its revealment the solutions of the Helmholtz equation in free space were matched at the interface of the crystal to those of the Dirac equation, which is applicable inside the photonic crystal in the vicinity of the Dirac point. Furthermore, conservation of $|\vec k_{||}|$, that is, Snell's law, and of the flux through the boundary had to be taken into account. In Fig.~\ref{fig5a} we show the data of the middle panel of Fig.~\ref{fig5} in a double-logarithmic plot together with the prediction for their average behavior (full line) obtained by evaluating Eqs.~(4)-(6) of Ref.~\cite{Sepkhanov2008}. Here, the range of angles $\pm\Delta\theta$ of the propagation direction with respect to the $\Gamma{\rm K}$ direction of the waves impinging the crystal is $\Delta\theta=14^\circ$. This yields for the range $\pm{\it\Delta}$ of the displacement of the wave vector $\vec k$ from the {\it K} point at {\bf K}, that is, of $|\vec k_{||}|$, the value ${\it\Delta}=\Delta\theta\pi f_D/c\simeq (\pi/3)a^{-1}$. For $\log_{10}(L/a)\gtrsim 0.7$. Hence, since the thickness of a layer equals $a/2$ (see Fig.~\ref{fig1}), at least $10$ layers are needed, in order that the average transmission $|S_{ba}|^2$ through the crystal shows the $1/L$ scaling predicted for $L/a > 1/(a{\it\Delta})=3/\pi$. It has been observed before in two-dimensional sonic crystals~\cite{Zhang2008} and in photonic crystals consisting of dielectric rods~\cite{Zandbergen2010}. There, in addition, oscillations of the transmission were found for an arrangement of the crystal with respect to the wave source as in Fig.~\ref{fig3} and attributed to the scattering between the valleys at ${\bf K}$ and ${\bf K^\prime}$. Oscillations are also visible in the double-logarithmic plot of the transmission shown in Fig.~\ref{fig5a}. They are, however, not as pronounced as those presented in Refs.~\cite{Zhang2008,Zandbergen2010}. In distinction to these experiments, successively two rows, instead of one row of cylinders were removed for the measurement of the data presented in Figs.~\ref{fig5} and~\ref{fig5a}. Nevertheless, the period of the oscillations is well reproduced by the equation~\cite{Zandbergen2010} (dashed line)
\begin{equation}
\vert S_{ba}\vert^2=Aa/L\left(1-\exp[-L{\it \Delta}]\right)\left([ 1+g+g\cos[\vert{\bf K}\vert L-\Phi]\right)\, ,
\label{eq:1}
\end{equation}
where $A$, $g$ and $\Phi$ are fitting parameters. The fit of Eq.~(\ref{eq:1}) to the date yielded $A=0.015$, $g=-0.55$ and $\Phi\approx\pi/3$.  

\subsection{Edge states in a rectangular Dirac billiard \label{Edge}}
A motivation for the experiments presented in this section were the results obtained for the spectral properties of graphene flakes~\cite{Nakada1996,Castro2009,Wurm2011,Ponomarenko2008,Wurm2009} and Ref.~\cite{Berry1987}. There Berry and Mondragon considered the Dirac equation for massless spin one-half particles confined to a two-dimensional domain by introducing an infinite-mass term along its boundary, i.e., a so-called neutrino billiard. Close to the Dirac frequency the propagation of electromagnetic waves in a photonic crystal can be effectively described by the Dirac equation. Thus, enclosing a microwave photonic crystal by a metallic frame, where the electric field strength obeys the Dirichlet boundary condition and the outgoing current vanishes, corresponds to introducing an infinite-mass term in the associated Dirac equation.  Accordingly, below $f=f_{\rm max}$ this provides an experimental setup for the investigation of properties of relativistic quantum particles put inside an infinitely high potential barrier. The reflection of electromagnetic waves at the walls of the metallic box leads to an excitation of states in both the valleys at ${\bf K}$ and ${\bf K^\prime}$ of the associated band structure. Their coupling depends on the edge structure. Therefore, in distinction to neutrino billiards, such systems are governed by two Dirac equations, which are coupled depending on the shape of the metallic box enclosing the crystal~\cite{Brey2006,Wurm2009}. They are generally called Dirac billiards~\cite{Libisch2009,Ponomarenko2008}.  
\begin{figure}[t]
{\includegraphics[width=\linewidth]{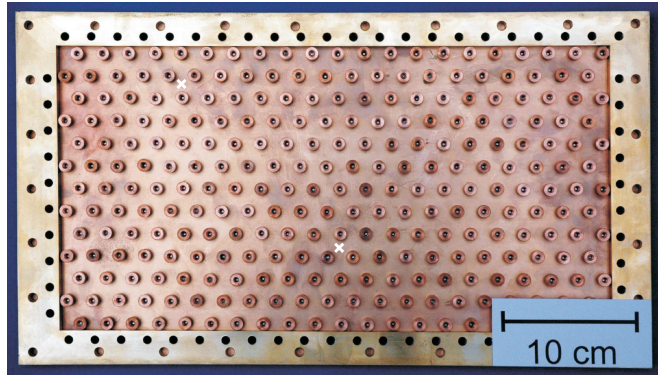}}
\caption{(Color online) Photograph of the Dirac billiard with the top plate removed. The photonic crystal with a triangular lattice geomatrey is enclosed by a rectangular brass frame with side lengths of, respectively, $218$~mm and $420$~mm. It consisted of $273$ copper cylinders of radius $R=5$~mm and height $h=8$~mm, had a lattice constant $a=4R$ and was squeezed between two copper plates. The two white crosses mark the positions of the antennas.} 
\label{fig6}
\end{figure}
Figure~\ref{fig6} shows a photograph of the experimental realization of a rectangular Dirac billiard, where the top plate has been removed. The photonic crystal had the same lattice constant $a=4R$ and consisted of $273$ copper cylinders with the same radius $R=5$~mm and height $h=8$~mm as in the scattering experiment described in Sec.~\ref{Extrem}. It is surrounded by a brass frame with side lengths of, respectively, $218$~mm and $420$~mm. In the experiments all cylinders and the frame were squeezed with screws between the top and bottom plates to ensure an optimal electrical contact. To measure transmission spectra wire antennas of $1$~mm diameter were lowered into the resonator through 3~mm wide drillings in the top plate. They reached only $2$~mm into the interior of the cavity to minimize the disturbance of the electric field modes excited there. The positions of the antennas are marked by white crosses in Fig.~\ref{fig6}.

\begin{figure}[!t]
{\includegraphics[width=\linewidth]{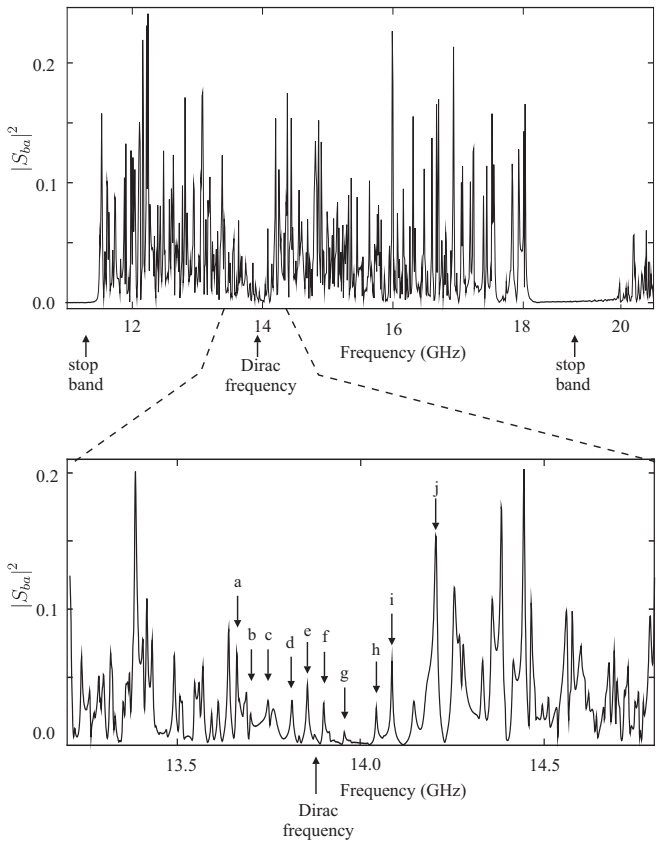}}
\caption{Top panel: transmission spectrum measured between two wire antennas. Bottom panel: magnification of the spectrum in a region around the Dirac frequency. The arrows mark the resonance frequencies for which the measured electric field intensity distributions are  shown on Fig.~\ref{fig8}. The labels of the resonances refer to the panels in Fig.~\ref{fig8}.} 
\label{fig7}
\end{figure}
The top panel of Fig.~\ref{fig7} shows a measured transmission spectrum of the Dirac billiard. The positions of the resonance maxima yield the eigenvalues of the Dirac billiard. The stop bands below $11.8$~GHz and between $18$ and $20$~GHz are well pronounced. As in the scattering experiment described in Sec.~\ref{Extrem} we observe a dip around the Dirac frequency $f_D=13.797$~GHz, which is marked with an arrow. The bottom panel of Fig.~\ref{fig7} shows a magnified section of the transmission spectrum around the Dirac point. The spectrum contains sharp resonances and many overlapping ones. The latter render the extraction of the complete sequence of resonance frequencies impossible. However, complete sequences of the resonance frequencies are needed for the study of their fluctuation properties. This corroborates the necessity of the measurement of the spectra at superconducting conditions~\cite{Graf1992,Dembowski2005} in order to reduce dissipation of microwave power in the walls of the metallic cylinders and the resonator enclosing it, yielding considerably diminished widths of the resonances. Still, in the vicinity of the Dirac frequency the density of the resonances is low such that they can be resolved. In fact, there the local density of states tends to zero linearly with the distance $|f-f_D|$ of the excitation frequency $f$ from the Dirac frequency $f_D$~\cite{Wallace1947,Castro2009,Slonczewski1958,Bittner2010}. Note that only in this frequency range the properties of the microwave Dirac billiard are well described by the Dirac equation, that is, only there it can simulate a Dirac billiard. The associated frequency range has been determined in Ref.~\cite{Bittner2010} using the relation between the local density of states and reflection spectra measured with one antenna placed in the interior of the photonic crystal. It extends from $f\simeq 13.3$~GHz to $f\simeq 14.3$~GHz. In this range of separated resonances the corresponding wave functions could be measured and their properties, such as the occurence of edge states predicted for graphene billiards~\cite{Wurm2011}, were investigated.

The wave functions of a quantum billiard are directly related to the electric field strength distributions in the corresponding microwave billiard at the resonance frequencies~\cite{StoeckmannBuch2000}. Thus, their intensities can be determined experimentally by measuring the electric field intensity distributions. This was done with the help of the so-called perturbation body method~\cite{Sridhar1991,Gokirmak1998,Dembowski1999}, which is based on the Maier-Slater theorem~\cite{Maier1952a}. A small perturbation body is inserted into the cavity. This induces a shift of the resonance frequencies the size of which depends on the electric and magnetic field strengths at the position of the perturbation body. Since we are only interested in the intensity distribution of the wave functions of the corresponding Dirac billiard, we used a perturbation body made from magnetic rubber~\cite{Bogomol2006}. In order to obtain the spatial distribution of the electric field strength the perturbation body was moved in the closed cavity in 2~mm steps between the $13$ rows of cylinders which are parallel to the long sides of the billiard. This was done with an external guiding magnet, which was placed on a computer controlled positioning unit. For each position the transmission between the two antennas marked in Fig.~\ref{fig6} was measured.
\begin{figure}[!t]
{\includegraphics[width=\linewidth]{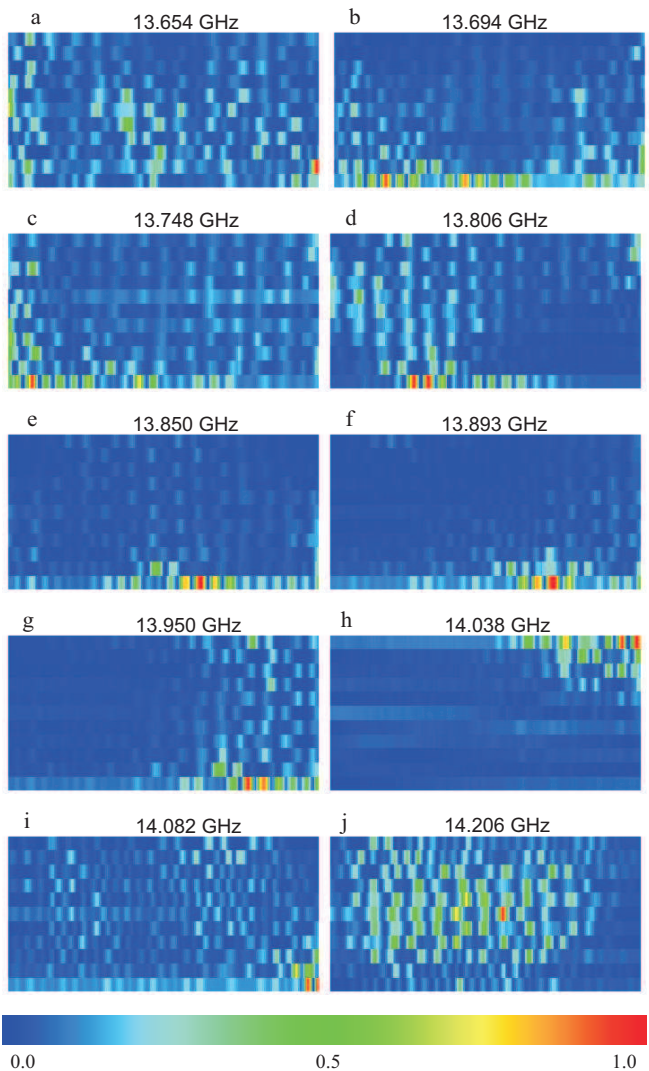}}
\caption{(Color online) The electric field intensity distributions measured at the resonance frequencies in Fig.~\ref{fig7}. These are marked with arrows and the labels ranging from a$\to$j. The blue (darkest) color corresponds to the lowest, the red (brightest) to the highest intensity.} 
\label{fig8}
\end{figure}

Figure~\ref{fig8} shows exemplarily ten electric field intensity distributions measured at resonance frequencies around the Dirac frequency. The labels of the panels correspond to the associated resonances in Fig.~\ref{fig7}. The intensity patterns shown in panels~a and j are extended over the whole area of the microwave billiard. The structure of these intensity distributions resembles that of modes of the corresponding empty rectangular quantum billiard with one excitation parallel to the short billiard edge and several parallel to the long one. They are similar to the wave functions shown in Fig. 3~(b) of Ref.~\cite{Libisch2009}. The intensity distributions shown in panels~c~-~i are localized close to a long edge of the rectangular billiard, i.e., there the intensity is highest and it is much smaller in the interior, especially in panels~e~-~h. At the long sides the photonic crystal terminates with straight edges, at the short sides with zigzag edges. These are close to, respectively, the zigzag edges and the armchair edges in the corresponding graphene billiard defined by the void structure of the photonic crystal. An example is shown in Fig.~\ref{fig1}, where the two sublattices are depicted as, respectively, red and blue dots. States which are localized near a short edge of the microwave billiard were not observed~\cite{Kuhl2010a}, whereas altogether $16$ corresponding to a sequence of resonances around the Dirac frequency were found to be localized near a long edge. This is in good agreement with the results for the associated rectangular graphene billiard~\cite{Wurm2011}. There, $14$ consecutive states around the Dirac point, which are all localized along a zigzag edge, so-called edge states, are predicted. Note that the boundary conditions of the microwave billiard and the corresponding graphene billiard differ. For the latter the wave functions vanish on one sublattice at a zigzag edge, and on both at an armchair edge, whereas for the former they are zero along the side walls, which are arranged slightly apart from the edges of the graphene billiard (see Fig.~\ref{fig1}). 
In spite of these differences both systems exhibit edge states corresponding to consecutive resonance frequencies around the Dirac frequency, which all are localized near a zigzag edge. They remind of the edge states observed experimentally~\cite{Kobayashi2005,Kobayashi2006,Tao2011} in graphene flakes, where electrons are confined by a potential barrier which, in distinction to that along the boundary of the Dirac billiard, is of finite height, and recently also in a magnetic photonic crystal~\cite{Poo2011}.  

\section{Summary}
In summary, we have observed experimentally the anomalous $1/L$ scaling behavior of the transmission of electromagnetic waves through a photonic crystal, which were excited at the Dirac frequency and sent onto the crystal in directions close to the $\Gamma {\rm K}$ direction, that is, near a conically shaped singularity of the photonic band structure. Furthermore, for the first time we implemented an experimental setup representing a microwave analog of a Dirac billiard, that is a relativistic quantum billiard. We have observed experimentally in the vicinity of the Dirac point states localized at edges of the Dirac billiard, which are similar to the edge states occurring at zigzag edges in graphene flakes. These similarities between the properties of waves propagating through a photonic crystal and those of graphene corroborate experimentally that not only the linear dispersion relation but also the unusual properties of graphene associated with it are entirely due to the symmetry properties of its lattice. Furthermore, this demonstrates that microwave billiards containing a photonic crystal provide an experimental setup for the investigation of properties of Dirac billiards of different shapes and with varying boundary conditions. These systems are particularly suited for the investigation of the spectral properties of Dirac billiards since the measurements can be performed with superconducting microwave billiards at liquid helium temperature. The aim is the study of the effect of the shape of a Dirac billiard and the edge structure of the enclosed photonic crystal on its spectral properties. Another interesting aspect concerning properties of the wave functions is the scarring phenomenon which has been seen numerically in graphene flakes with the shape of a stadium~\cite{Huang2009}.

\section{Acknowledgements}
\par The authors thank E. Bogomolny and T. Tudorovskiy for stimulating discussions, and the LPTMS in Orsay for its hospitality. This work has been supported by the DFG within the SFB 634. 


\begin{thebibliography}{42}
\expandafter\ifx\csname natexlab\endcsname\relax\def\natexlab#1{#1}\fi
\expandafter\ifx\csname bibnamefont\endcsname\relax
  \def\bibnamefont#1{#1}\fi
\expandafter\ifx\csname bibfnamefont\endcsname\relax
  \def\bibfnamefont#1{#1}\fi
\expandafter\ifx\csname citenamefont\endcsname\relax
  \def\citenamefont#1{#1}\fi
\expandafter\ifx\csname url\endcsname\relax
  \def\url#1{\texttt{#1}}\fi
\expandafter\ifx\csname urlprefix\endcsname\relax\def\urlprefix{URL }\fi
\providecommand{\bibinfo}[2]{#2}
\providecommand{\eprint}[2][]{\url{#2}}

\bibitem[{\citenamefont{Novoselov et~al.}(2004)\citenamefont{Novoselov, Geim,
  Morozov, Jiang, Zhang, Dubonos, Grigorieva, and Firsov}}]{Novoselov2004}
\bibinfo{author}{\bibfnamefont{K.~S.} \bibnamefont{Novoselov}},
  \bibinfo{author}{\bibfnamefont{A.~K.} \bibnamefont{Geim}},
  \bibinfo{author}{\bibfnamefont{S.~V.} \bibnamefont{Morozov}},
  \bibinfo{author}{\bibfnamefont{D.}~\bibnamefont{Jiang}},
  \bibinfo{author}{\bibfnamefont{Y.}~\bibnamefont{Zhang}},
  \bibinfo{author}{\bibfnamefont{S.~V.} \bibnamefont{Dubonos}},
  \bibinfo{author}{\bibfnamefont{I.~V.} \bibnamefont{Grigorieva}},
  \bibnamefont{and} \bibinfo{author}{\bibfnamefont{A.~A.}
  \bibnamefont{Firsov}}, \bibinfo{journal}{Science}
  \textbf{\bibinfo{volume}{306}}, \bibinfo{pages}{666} (\bibinfo{year}{2004}).

\bibitem[{\citenamefont{Wallace}(1947)}]{Wallace1947}
\bibinfo{author}{\bibfnamefont{P.~R.} \bibnamefont{Wallace}},
  \bibinfo{journal}{Phys. Rev.} \textbf{\bibinfo{volume}{71}},
  \bibinfo{pages}{622} (\bibinfo{year}{1947}).

\bibitem[{\citenamefont{Boehm et~al.}(1994)\citenamefont{Boehm, Setton, and
  Stumpp}}]{Boehm1994}
\bibinfo{author}{\bibfnamefont{H.-P.}~\bibnamefont{Boehm}},
  \bibinfo{author}{\bibfnamefont{R.}~\bibnamefont{Setton}}, \bibnamefont{and}
  \bibinfo{author}{\bibfnamefont{E.}~\bibnamefont{Stumpp}},
  \bibinfo{journal}{Pure Appl. Chem.} \textbf{\bibinfo{volume}{66}},
  \bibinfo{pages}{1893} (\bibinfo{year}{1994}).

\bibitem[{\citenamefont{Castro~Neto et~al.}(2009)\citenamefont{Castro~Neto,
  Guinea, Peres, Novoselov, and Geim}}]{Castro2009}
\bibinfo{author}{\bibfnamefont{A.~H.} \bibnamefont{Castro~Neto}},
  \bibinfo{author}{\bibfnamefont{F.}~\bibnamefont{Guinea}},
  \bibinfo{author}{\bibfnamefont{N.~M.~R.} \bibnamefont{Peres}},
  \bibinfo{author}{\bibfnamefont{K.~S.} \bibnamefont{Novoselov}},
  \bibnamefont{and} \bibinfo{author}{\bibfnamefont{A.~K.} \bibnamefont{Geim}},
  \bibinfo{journal}{Rev. Mod. Phys.} \textbf{\bibinfo{volume}{81}},
  \bibinfo{pages}{109} (\bibinfo{year}{2009}).

\bibitem[{\citenamefont{Beenakker}(2008)}]{Beenakker2008}
\bibinfo{author}{\bibfnamefont{C.~W.~J.} \bibnamefont{Beenakker}},
  \bibinfo{journal}{Rev. Mod. Phys.} \textbf{\bibinfo{volume}{80}},
  \bibinfo{pages}{1337} (\bibinfo{year}{2008}).

\bibitem[{\citenamefont{Slonczewski and Weiss}(1958)}]{Slonczewski1958}
\bibinfo{author}{\bibfnamefont{J.~C.} \bibnamefont{Slonczewski}}
  \bibnamefont{and} \bibinfo{author}{\bibfnamefont{P.~R.} \bibnamefont{Weiss}},
  \bibinfo{journal}{Phys. Rev.} \textbf{\bibinfo{volume}{109}},
  \bibinfo{pages}{272} (\bibinfo{year}{1958}).

\bibitem[{\citenamefont{Joannopoulos et~al.}(2008)\citenamefont{Joannopoulos,
  Meade, and Winn}}]{Joannopoulos2008}
\bibinfo{author}{\bibfnamefont{S.}~\bibnamefont{Joannopoulos},
  \bibfnamefont{J.~D.~Johnson}},
  \bibinfo{author}{\bibfnamefont{R.}~\bibnamefont{Meade}}, \bibnamefont{and}
  \bibinfo{author}{\bibfnamefont{J.}~\bibnamefont{Winn}},
  \emph{\bibinfo{title}{Photonic Crystals. Molding the Flow of Light}}
  (\bibinfo{publisher}{Princeton University Press, Princeton and Oxford},
  \bibinfo{year}{2008}), \bibinfo{edition}{2nd} ed.

\bibitem[{\citenamefont{Raghu and Haldane}(2008)}]{Raghu2008}
\bibinfo{author}{\bibfnamefont{S.}~\bibnamefont{Raghu}} \bibnamefont{and}
  \bibinfo{author}{\bibfnamefont{F.~D.~M.} \bibnamefont{Haldane}},
  \bibinfo{journal}{Phys. Rev. A} \textbf{\bibinfo{volume}{78}},
  \bibinfo{pages}{033834} (\bibinfo{year}{2008}).

\bibitem[{\citenamefont{Smirnova et~al.}(2002)\citenamefont{Smirnova, Chen,
  Shapiro, Sirigiri, and Temkin}}]{Smirnova2002}
\bibinfo{author}{\bibfnamefont{E.~I.} \bibnamefont{Smirnova}},
  \bibinfo{author}{\bibfnamefont{C.}~\bibnamefont{Chen}},
  \bibinfo{author}{\bibfnamefont{M.~A.} \bibnamefont{Shapiro}},
  \bibinfo{author}{\bibfnamefont{J.~R.} \bibnamefont{Sirigiri}},
  \bibnamefont{and} \bibinfo{author}{\bibfnamefont{R.~J.}
  \bibnamefont{Temkin}}, \bibinfo{journal}{J. Appl. Phys.}
  \textbf{\bibinfo{volume}{91}}, \bibinfo{pages}{960} (\bibinfo{year}{2002}).

\bibitem[{\citenamefont{Katsnelson}(2007)}]{Katsnelson2007}
\bibinfo{author}{\bibfnamefont{M.}~\bibnamefont{Katsnelson}},
  \bibinfo{journal}{Mater. Today} \textbf{\bibinfo{volume}{10}},
  \bibinfo{pages}{20} (\bibinfo{year}{2007}).

\bibitem[{\citenamefont{Novoselov et~al.}(2005)\citenamefont{Novoselov, Geim,
  Morozov, Jiang, Katsnelson, Grigorieva, Dubonos, and Firsov}}]{Novoselov2005}
\bibinfo{author}{\bibfnamefont{K.~S.} \bibnamefont{Novoselov}},
  \bibinfo{author}{\bibfnamefont{A.~K.} \bibnamefont{Geim}},
  \bibinfo{author}{\bibfnamefont{S.~V.} \bibnamefont{Morozov}},
  \bibinfo{author}{\bibfnamefont{D.}~\bibnamefont{Jiang}},
  \bibinfo{author}{\bibfnamefont{M.~I.} \bibnamefont{Katsnelson}},
  \bibinfo{author}{\bibfnamefont{I.}~\bibnamefont{Grigorieva}},
  \bibinfo{author}{\bibfnamefont{S.~V.} \bibnamefont{Dubonos}},
  \bibnamefont{and} \bibinfo{author}{\bibfnamefont{A.~A.}
  \bibnamefont{Firsov}}, \bibinfo{journal}{Nature}
  \textbf{\bibinfo{volume}{438}}, \bibinfo{pages}{197} (\bibinfo{year}{2005}).

\bibitem[{\citenamefont{Gaspard and Rice}(1989)}]{Gaspard1989}
\bibinfo{author}{\bibfnamefont{P.}~\bibnamefont{Gaspard}} \bibnamefont{and}
  \bibinfo{author}{\bibfnamefont{S.~A.} \bibnamefont{Rice}},
  \bibinfo{journal}{J. Chem. Phys.} \textbf{\bibinfo{volume}{90}},
  \bibinfo{pages}{2255} (\bibinfo{year}{1989}).

\bibitem[{\citenamefont{Brey and Fertig}(2006)}]{Brey2006}
\bibinfo{author}{\bibfnamefont{L.}~\bibnamefont{Brey}} \bibnamefont{and}
  \bibinfo{author}{\bibfnamefont{H.~A.} \bibnamefont{Fertig}},
  \bibinfo{journal}{Phys. Rev. B} \textbf{\bibinfo{volume}{73}},
  \bibinfo{eid}{235411} (\bibinfo{year}{2006}).

\bibitem[{\citenamefont{Ando}(2005)}]{Ando2005}
\bibinfo{author}{\bibfnamefont{T.}~\bibnamefont{Ando}}, \bibinfo{journal}{J.
  Phys. Soc. Jap.} \textbf{\bibinfo{volume}{74}}, \bibinfo{pages}{777}
  (\bibinfo{year}{2005}).

\bibitem[{\citenamefont{Zandbergen and de~Dood}(2010)}]{Zandbergen2010}
\bibinfo{author}{\bibfnamefont{S.~R.} \bibnamefont{Zandbergen}}
  \bibnamefont{and} \bibinfo{author}{\bibfnamefont{M.~J.~A.}
  \bibnamefont{de~Dood}}, \bibinfo{journal}{Phys. Rev. Lett.}
  \textbf{\bibinfo{volume}{104}}, \bibinfo{pages}{043903}
  (\bibinfo{year}{2010}).

\bibitem[{\citenamefont{Kuhl et~al.}(2010)\citenamefont{Kuhl, Barkhofen,
  Tudorovskiy, St\"ockmann, Hossain, de~Forges~de Parny, and
  Mortessagne}}]{Kuhl2010a}
\bibinfo{author}{\bibfnamefont{U.}~\bibnamefont{Kuhl}},
  \bibinfo{author}{\bibfnamefont{S.}~\bibnamefont{Barkhofen}},
  \bibinfo{author}{\bibfnamefont{T.}~\bibnamefont{Tudorovskiy}},
  \bibinfo{author}{\bibfnamefont{H.-J.} \bibnamefont{St\"ockmann}},
  \bibinfo{author}{\bibfnamefont{T.}~\bibnamefont{Hossain}},
  \bibinfo{author}{\bibfnamefont{L.}~\bibnamefont{de~Forges~de Parny}},
  \bibnamefont{and}
  \bibinfo{author}{\bibfnamefont{F.}~\bibnamefont{Mortessagne}},
  \bibinfo{journal}{Phys. Rev. B} \textbf{\bibinfo{volume}{82}},
  \bibinfo{pages}{094308} (\bibinfo{year}{2010}).

\bibitem[{\citenamefont{Bittner et~al.}(2010)\citenamefont{Bittner, Dietz,
  Miski-Oglu, Oria~Iriarte, Richter, and Sch\"afer}}]{Bittner2010}
\bibinfo{author}{\bibfnamefont{S.}~\bibnamefont{Bittner}},
  \bibinfo{author}{\bibfnamefont{B.}~\bibnamefont{Dietz}},
  \bibinfo{author}{\bibfnamefont{M.}~\bibnamefont{Miski-Oglu}},
  \bibinfo{author}{\bibfnamefont{P.}~\bibnamefont{Oria~Iriarte}},
  \bibinfo{author}{\bibfnamefont{A.}~\bibnamefont{Richter}}, \bibnamefont{and}
  \bibinfo{author}{\bibfnamefont{F.}~\bibnamefont{Sch\"afer}},
  \bibinfo{journal}{Phys. Rev. B} \textbf{\bibinfo{volume}{82}},
  \bibinfo{pages}{014301} (\bibinfo{year}{2010}).

\bibitem[{\citenamefont{Zhang and Liu}(2008)}]{Zhang2008}
\bibinfo{author}{\bibfnamefont{X.}~\bibnamefont{Zhang}} \bibnamefont{and}
  \bibinfo{author}{\bibfnamefont{Z.}~\bibnamefont{Liu}},
  \bibinfo{journal}{Phys. Rev. Lett.} \textbf{\bibinfo{volume}{101}},
  \bibinfo{eid}{264303} (\bibinfo{year}{2008}).

\bibitem[{\citenamefont{Sepkhanov et~al.}(2007)\citenamefont{Sepkhanov,
  Bazaliy, and Beenakker}}]{Sepkhanov2007}
\bibinfo{author}{\bibfnamefont{R.~A.} \bibnamefont{Sepkhanov}},
  \bibinfo{author}{\bibfnamefont{Y.~B.} \bibnamefont{Bazaliy}},
  \bibnamefont{and} \bibinfo{author}{\bibfnamefont{C.~W.~J.}
  \bibnamefont{Beenakker}}, \bibinfo{journal}{Phys. Rev. A}
  \textbf{\bibinfo{volume}{75}}, \bibinfo{eid}{063813} (\bibinfo{year}{2007}).

\bibitem[{\citenamefont{Sepkhanov et~al.}(2008)\citenamefont{Sepkhanov 
  and Beenakker}}]{Sepkhanov2008}
\bibinfo{author}{\bibfnamefont{R.~A.} \bibnamefont{Sepkhanov}}
  \bibnamefont{and} \bibinfo{author}{\bibfnamefont{C.~W.~J.}
  \bibnamefont{Beenakker}}, \bibinfo{journal}{Opt. Commun.}
  \textbf{\bibinfo{volume}{281}}, \bibinfo{eid}{5267} (\bibinfo{year}{2008}).

\bibitem[{\citenamefont{Tworzyd\l{}o et~al.}(2006)\citenamefont{Tworzyd\l{}o,
  Trauzettel, Titov, Rycerz, and Beenakker}}]{Tworzydlo2006}
\bibinfo{author}{\bibfnamefont{J.}~\bibnamefont{Tworzyd\l{}o}},
  \bibinfo{author}{\bibfnamefont{B.}~\bibnamefont{Trauzettel}},
  \bibinfo{author}{\bibfnamefont{M.}~\bibnamefont{Titov}},
  \bibinfo{author}{\bibfnamefont{A.}~\bibnamefont{Rycerz}}, \bibnamefont{and}
  \bibinfo{author}{\bibfnamefont{C.~W.~J.} \bibnamefont{Beenakker}},
  \bibinfo{journal}{Phys. Rev. Lett.} \textbf{\bibinfo{volume}{96}},
  \bibinfo{pages}{246802} (\bibinfo{year}{2006}).

\bibitem[{\citenamefont{Katsnelson}(2006)}]{Katsnelson2006a}
\bibinfo{author}{\bibfnamefont{M.~I.} \bibnamefont{Katsnelson}},
  \bibinfo{journal}{Eur. Phys. J. B} \textbf{\bibinfo{volume}{51}},
  \bibinfo{pages}{157} (\bibinfo{year}{2006}).

\bibitem[{\citenamefont{Nakada et~al.}(1996)\citenamefont{Nakada, Fujita,
  Dresselhaus, and Dresselhaus}}]{Nakada1996}
\bibinfo{author}{\bibfnamefont{K.}~\bibnamefont{Nakada}},
  \bibinfo{author}{\bibfnamefont{M.}~\bibnamefont{Fujita}},
  \bibinfo{author}{\bibfnamefont{G.}~\bibnamefont{Dresselhaus}},
  \bibnamefont{and} \bibinfo{author}{\bibfnamefont{M.~S.}
  \bibnamefont{Dresselhaus}}, \bibinfo{journal}{Phys. Rev. B}
  \textbf{\bibinfo{volume}{54}}, \bibinfo{pages}{17954} (\bibinfo{year}{1996}).

\bibitem[{\citenamefont{Wurm et~al.}(2011)\citenamefont{Wurm, Richter, and
  Adagideli}}]{Wurm2011}
\bibinfo{author}{\bibfnamefont{J.}~\bibnamefont{Wurm}},
  \bibinfo{author}{\bibfnamefont{K.}~\bibnamefont{Richter}}, \bibnamefont{and}
  \bibinfo{author}{\bibfnamefont{\.I.} \bibnamefont{Adagideli}},
  \bibinfo{journal}{Phys. Rev. B} \textbf{\bibinfo{volume}{84}},
  \bibinfo{pages}{075468} (\bibinfo{year}{2011}).

\bibitem[{\citenamefont{Ponomarenko et~al.}(2008)\citenamefont{Ponomarenko,
  Schedin, Katsnelson, Yang, Hill, Novoselov, and Geim}}]{Ponomarenko2008}
\bibinfo{author}{\bibfnamefont{L.~A.} \bibnamefont{Ponomarenko}},
  \bibinfo{author}{\bibfnamefont{F.}~\bibnamefont{Schedin}},
  \bibinfo{author}{\bibfnamefont{M.~I.} \bibnamefont{Katsnelson}},
  \bibinfo{author}{\bibfnamefont{R.}~\bibnamefont{Yang}},
  \bibinfo{author}{\bibfnamefont{E.~W.} \bibnamefont{Hill}},
  \bibinfo{author}{\bibfnamefont{K.~S.} \bibnamefont{Novoselov}},
  \bibnamefont{and} \bibinfo{author}{\bibfnamefont{A.~K.} \bibnamefont{Geim}},
  \bibinfo{journal}{Science} \textbf{\bibinfo{volume}{320}},
  \bibinfo{pages}{5874} (\bibinfo{year}{2008}).

\bibitem[{\citenamefont{Wurm et~al.}(2009)\citenamefont{Wurm, Rycerz,
  Adagideli, Wimmer, Richter, and Baranger}}]{Wurm2009}
\bibinfo{author}{\bibfnamefont{J.}~\bibnamefont{Wurm}},
  \bibinfo{author}{\bibfnamefont{A.}~\bibnamefont{Rycerz}},
  \bibinfo{author}{\bibfnamefont{{\.I}.}~\bibnamefont{Adagideli}},
  \bibinfo{author}{\bibfnamefont{M.}~\bibnamefont{Wimmer}},
  \bibinfo{author}{\bibfnamefont{K.}~\bibnamefont{Richter}}, \bibnamefont{and}
  \bibinfo{author}{\bibfnamefont{H.~U.} \bibnamefont{Baranger}},
  \bibinfo{journal}{Phys. Rev. Lett.} \textbf{\bibinfo{volume}{102}},
  \bibinfo{pages}{056806} (\bibinfo{year}{2009}).

\bibitem[{\citenamefont{Libisch et~al.}(2009)\citenamefont{Libisch, Stampfer,
  and Burgd\"{o}rfer}}]{Libisch2009}
\bibinfo{author}{\bibfnamefont{F.}~\bibnamefont{Libisch}},
  \bibinfo{author}{\bibfnamefont{C.}~\bibnamefont{Stampfer}}, \bibnamefont{and}
  \bibinfo{author}{\bibfnamefont{J.}~\bibnamefont{Burgd\"{o}rfer}},
  \bibinfo{journal}{Phys. Rev. B} \textbf{\bibinfo{volume}{79}},
  \bibinfo{eid}{115423} (\bibinfo{year}{2009}).

\bibitem[{\citenamefont{Kobayashi et~al.}(2005)\citenamefont{Kobayashi, Fukui,
Enoki, Kusakabe and Kaburagi}}]{Kobayashi2005}
\bibinfo{author}{\bibfnamefont{Y.}~\bibnamefont{Kobayashi}},
\bibinfo{author}{\bibfnamefont{K. I.}~\bibnamefont{Fukui}}, 
\bibinfo{author}{\bibfnamefont{T.}~\bibnamefont{Enoki}},
\bibinfo{author}{\bibfnamefont{K.}~\bibnamefont{Kusakabe}}, 
\bibnamefont{and}
  \bibinfo{author}{\bibfnamefont{Y.}~\bibnamefont{Kaburagi}},
  \bibinfo{journal}{Phys. Rev. B} \textbf{\bibinfo{volume}{71}},
  \bibinfo{eid}{193406} (\bibinfo{year}{2005}).

\bibitem[{\citenamefont{Kobayashi et~al.}(2006)\citenamefont{Kobayashi, Fukui,
Enoki and Kusakabe}}]{Kobayashi2006}
\bibinfo{author}{\bibfnamefont{Y.}~\bibnamefont{Kobayashi}},
\bibinfo{author}{\bibfnamefont{K. I.}~\bibnamefont{Fukui}}, 
\bibinfo{author}{\bibfnamefont{T.}~\bibnamefont{Enoki}},
\bibnamefont{and}
  \bibinfo{author}{\bibfnamefont{K.}~\bibnamefont{Kusakabe}}, 
  \bibinfo{journal}{Phys. Rev. B} \textbf{\bibinfo{volume}{73}},
  \bibinfo{eid}{125415} (\bibinfo{year}{2006}).

\bibitem[{\citenamefont{Tao et~al.}(2011)\citenamefont{Tao, Jiao,
Yazyev, Chen, Feng, Zhang, Capaz, Tour, Zettl, Louie, Dai and Crommie}}]{Tao2011}
\bibinfo{author}{\bibfnamefont{C.}~\bibnamefont{Tao}},
\bibinfo{author}{\bibfnamefont{L.}~\bibnamefont{Jiao}}, 
\bibinfo{author}{\bibfnamefont{O.}~\bibnamefont{Yazyev}},
\bibinfo{author}{\bibfnamefont{Y.-C.}~\bibnamefont{Chen}},
\bibinfo{author}{\bibfnamefont{J.}~\bibnamefont{Feng}}, 
\bibinfo{author}{\bibfnamefont{X.}~\bibnamefont{Zhang}},
\bibinfo{author}{\bibfnamefont{R.}~\bibnamefont{Capaz}},
\bibinfo{author}{\bibfnamefont{J. M.}~\bibnamefont{Tour}}, 
\bibinfo{author}{\bibfnamefont{A.}~\bibnamefont{Zettl}},
\bibinfo{author}{\bibfnamefont{S. G.}~\bibnamefont{Louie}},
\bibinfo{author}{\bibfnamefont{H.}~\bibnamefont{Dai}}, 
\bibnamefont{and}
  \bibinfo{author}{\bibfnamefont{M.}~\bibnamefont{Crommie}}, 
  \bibinfo{journal}{Nature Phys.} \textbf{\bibinfo{volume}{7}},
  \bibinfo{eid}{616} (\bibinfo{year}{2011}).

\bibitem[{\citenamefont{Akhmerov and Beenakker}(2008)}]{Akhmerov2008}
\bibinfo{author}{\bibfnamefont{A.~R.} \bibnamefont{Akhmerov}} \bibnamefont{and}
  \bibinfo{author}{\bibfnamefont{C.~W.~J.} \bibnamefont{Beenakker}},
  \bibinfo{journal}{Phys. Rev. B} \textbf{\bibinfo{volume}{77}},
  \bibinfo{pages}{085423} (\bibinfo{year}{2008}).

\bibitem[{\citenamefont{Sridhar}(1991)}]{Sridhar1991}
\bibinfo{author}{\bibfnamefont{S.}~\bibnamefont{Sridhar}},
  \bibinfo{journal}{Phys. Rev. Lett.} \textbf{\bibinfo{volume}{67}},
  \bibinfo{pages}{785} (\bibinfo{year}{1991}).

\bibitem[{\citenamefont{Richter}(1999)}]{Richter1999}
\bibinfo{author}{\bibfnamefont{A.}~\bibnamefont{Richter}}, in
  \emph{\bibinfo{booktitle}{Emerging Applications of Number Theory, \rm{The IMA
  Volumes in Mathematics and its Applications}}}, edited by
  \bibinfo{editor}{\bibfnamefont{D.~A.} \bibnamefont{Hejhal}},
  \bibinfo{editor}{\bibfnamefont{J.}~\bibnamefont{Friedmann}},
  \bibinfo{editor}{\bibfnamefont{M.~C.} \bibnamefont{Gutzwiller}},
  \bibnamefont{and} \bibinfo{editor}{\bibfnamefont{A.~M.}
  \bibnamefont{Odlyzko}} (\bibinfo{publisher}{Springer}, \bibinfo{address}{New
  York}, \bibinfo{year}{1999}), vol. \bibinfo{volume}{109}, p.
  \bibinfo{pages}{479}.

\bibitem[{\citenamefont{St{\"o}ckmann}(2000)}]{StoeckmannBuch2000}
\bibinfo{author}{\bibfnamefont{H.-J.} \bibnamefont{St{\"o}ckmann}},
  \emph{\bibinfo{title}{Quantum Chaos: An Introduction}}
  (\bibinfo{publisher}{Cambridge University Press},
  \bibinfo{address}{Cambridge}, \bibinfo{year}{2000}).

\bibitem[{\citenamefont{Dietz et~al.}(2007)\citenamefont{Dietz, Friedrich,
  Miski-Oglu, Richter, and Sch\"{a}fer}}]{Dietz2007b}
\bibinfo{author}{\bibfnamefont{B.}~\bibnamefont{Dietz}},
  \bibinfo{author}{\bibfnamefont{T.}~\bibnamefont{Friedrich}},
  \bibinfo{author}{\bibfnamefont{M.}~\bibnamefont{Miski-Oglu}},
  \bibinfo{author}{\bibfnamefont{A.}~\bibnamefont{Richter}}, \bibnamefont{and}
  \bibinfo{author}{\bibfnamefont{F.}~\bibnamefont{Sch\"{a}fer}},
  \bibinfo{journal}{Phys. Rev. E} \textbf{\bibinfo{volume}{75}},
  \bibinfo{pages}{035203} (\bibinfo{year}{2007}).

\bibitem[{\citenamefont{Dietz et~al.}(2010)\citenamefont{Dietz, Friedrich,
  Harney, Miski-Oglu, Richter, Sch{\"a}fer, and Weidenm{\"u}ller}}]{Dietz2010}
\bibinfo{author}{\bibfnamefont{B.}~\bibnamefont{Dietz}},
  \bibinfo{author}{\bibfnamefont{T.}~\bibnamefont{Friedrich}},
  \bibinfo{author}{\bibfnamefont{H.~L.} \bibnamefont{Harney}},
  \bibinfo{author}{\bibfnamefont{M.}~\bibnamefont{Miski-Oglu}},
  \bibinfo{author}{\bibfnamefont{A.}~\bibnamefont{Richter}},
  \bibinfo{author}{\bibfnamefont{F.}~\bibnamefont{Sch{\"a}fer}},
  \bibnamefont{and} \bibinfo{author}{\bibfnamefont{H.~A.}
  \bibnamefont{Weidenm{\"u}ller}}, \bibinfo{journal}{Phys. Rev. E}
  \textbf{\bibinfo{volume}{81}}, \bibinfo{pages}{036205}
  (\bibinfo{year}{2010}).

\bibitem[{\citenamefont{Kuhl et~al.}(2005)\citenamefont{Kuhl, St\"ockmann, and
  Weaver}}]{Kuhl2005a}
\bibinfo{author}{\bibfnamefont{U.}~\bibnamefont{Kuhl}},
  \bibinfo{author}{\bibfnamefont{H.-J.} \bibnamefont{St\"ockmann}},
  \bibnamefont{and} \bibinfo{author}{\bibfnamefont{R.}~\bibnamefont{Weaver}},
  \bibinfo{journal}{J. Phys. A} \textbf{\bibinfo{volume}{38}},
  \bibinfo{pages}{10433} (\bibinfo{year}{2005}).

\bibitem{RMP} G. E. Mitchell, A. Richter, and H. A. Weidenm{\"u}ller, Rev. Mod. Phys. {\bf 82}, 2845 (2010).

\bibitem[{\citenamefont{Wang et~al.}(2009)\citenamefont{Wang, Wang, Zhang, and
  Zhu}}]{Wang2009a}
\bibinfo{author}{\bibfnamefont{L.-G.} \bibnamefont{Wang}},
  \bibinfo{author}{\bibfnamefont{Z.-G.} \bibnamefont{Wang}},
  \bibinfo{author}{\bibfnamefont{J.-X.} \bibnamefont{Zhang}}, \bibnamefont{and}
  \bibinfo{author}{\bibfnamefont{S.-Y.} \bibnamefont{Zhu}},
  \bibinfo{journal}{Opt. Lett.} \textbf{\bibinfo{volume}{34}},
  \bibinfo{pages}{1510} (\bibinfo{year}{2009}).

\bibitem[{\citenamefont{Berry and Mondragon}(1987)}]{Berry1987}
\bibinfo{author}{\bibfnamefont{M.~V.} \bibnamefont{Berry}} \bibnamefont{and}
  \bibinfo{author}{\bibfnamefont{R.~J.} \bibnamefont{Mondragon}},
  \bibinfo{journal}{Proc. R. Soc. London A} \textbf{\bibinfo{volume}{412}},
  \bibinfo{pages}{53} (\bibinfo{year}{1987}).

\bibitem[{\citenamefont{Gr\"af et~al.}(1992)\citenamefont{Gr\"af, Harney,
  Lengeler, Lewenkopf, Rangacharyulu, Richter, Schardt, and
  Weidenm\"uller}}]{Graf1992}
\bibinfo{author}{\bibfnamefont{H.-D.} \bibnamefont{Gr\"af}},
  \bibinfo{author}{\bibfnamefont{H.~L.} \bibnamefont{Harney}},
  \bibinfo{author}{\bibfnamefont{H.}~\bibnamefont{Lengeler}},
  \bibinfo{author}{\bibfnamefont{C.~H.} \bibnamefont{Lewenkopf}},
  \bibinfo{author}{\bibfnamefont{C.}~\bibnamefont{Rangacharyulu}},
  \bibinfo{author}{\bibfnamefont{A.}~\bibnamefont{Richter}},
  \bibinfo{author}{\bibfnamefont{P.}~\bibnamefont{Schardt}}, \bibnamefont{and}
  \bibinfo{author}{\bibfnamefont{H.~A.} \bibnamefont{Weidenm\"uller}},
  \bibinfo{journal}{Phys. Rev. Lett.} \textbf{\bibinfo{volume}{69}},
  \bibinfo{pages}{1296} (\bibinfo{year}{1992}).

\bibitem[{\citenamefont{Dembowski et~al.}(2005)\citenamefont{Dembowski, Dietz,
  Friedrich, Gr\"{a}f, Harney, Heine, Miski-Oglu, and Richter}}]{Dembowski2005}
\bibinfo{author}{\bibfnamefont{C.}~\bibnamefont{Dembowski}},
  \bibinfo{author}{\bibfnamefont{B.}~\bibnamefont{Dietz}},
  \bibinfo{author}{\bibfnamefont{T.}~\bibnamefont{Friedrich}},
  \bibinfo{author}{\bibfnamefont{H.-D.} \bibnamefont{Gr\"{a}f}},
  \bibinfo{author}{\bibfnamefont{H.~L.} \bibnamefont{Harney}},
  \bibinfo{author}{\bibfnamefont{A.}~\bibnamefont{Heine}},
  \bibinfo{author}{\bibfnamefont{M.}~\bibnamefont{Miski-Oglu}},
  \bibnamefont{and} \bibinfo{author}{\bibfnamefont{A.}~\bibnamefont{Richter}},
  \bibinfo{journal}{Phys. Rev. E} \textbf{\bibinfo{volume}{71}},
  \bibinfo{eid}{046202} (\bibinfo{year}{2005}).

\bibitem[{\citenamefont{Gokirmak et~al.}(1998)\citenamefont{Gokirmak, Wu,
  Bridgewater, and Anlage}}]{Gokirmak1998}
\bibinfo{author}{\bibfnamefont{A.}~\bibnamefont{Gokirmak}},
  \bibinfo{author}{\bibfnamefont{D.-H.} \bibnamefont{Wu}},
  \bibinfo{author}{\bibfnamefont{J.~S.~A.} \bibnamefont{Bridgewater}},
  \bibnamefont{and} \bibinfo{author}{\bibfnamefont{S.~M.}
  \bibnamefont{Anlage}}, \bibinfo{journal}{Rev.\ Sci.\ Instrum.}
  \textbf{\bibinfo{volume}{69}}, \bibinfo{pages}{3410} (\bibinfo{year}{1998}).

\bibitem[{\citenamefont{Dembowski et~al.}(1999)\citenamefont{Dembowski, Gr\"af,
  Hofferbert, Rehfeld, Richter, and Weiland}}]{Dembowski1999}
\bibinfo{author}{\bibfnamefont{C.}~\bibnamefont{Dembowski}},
  \bibinfo{author}{\bibfnamefont{H.-D.} \bibnamefont{Gr\"af}},
  \bibinfo{author}{\bibfnamefont{R.}~\bibnamefont{Hofferbert}},
  \bibinfo{author}{\bibfnamefont{H.}~\bibnamefont{Rehfeld}},
  \bibinfo{author}{\bibfnamefont{A.}~\bibnamefont{Richter}}, \bibnamefont{and}
  \bibinfo{author}{\bibfnamefont{T.}~\bibnamefont{Weiland}},
  \bibinfo{journal}{Phys. Rev. E} \textbf{\bibinfo{volume}{60}},
  \bibinfo{pages}{3942} (\bibinfo{year}{1999}).

\bibitem[{\citenamefont{Maier and Slater}(1952)}]{Maier1952a}
\bibinfo{author}{\bibfnamefont{L.~C.} \bibnamefont{Maier}} \bibnamefont{and}
  \bibinfo{author}{\bibfnamefont{J.~C.} \bibnamefont{Slater}},
  \bibinfo{journal}{J. Appl. Phys.} \textbf{\bibinfo{volume}{23}},
  \bibinfo{pages}{68} (\bibinfo{year}{1952}).

\bibitem[{\citenamefont{Bogomolny et~al.}(2006)\citenamefont{Bogomolny, Dietz,
  Friedrich, Miski-Oglu, Richter, Sch\"{a}fer, and Schmit}}]{Bogomol2006}
\bibinfo{author}{\bibfnamefont{E.}~\bibnamefont{Bogomolny}},
  \bibinfo{author}{\bibfnamefont{B.}~\bibnamefont{Dietz}},
  \bibinfo{author}{\bibfnamefont{T.}~\bibnamefont{Friedrich}},
  \bibinfo{author}{\bibfnamefont{M.}~\bibnamefont{Miski-Oglu}},
  \bibinfo{author}{\bibfnamefont{A.}~\bibnamefont{Richter}},
  \bibinfo{author}{\bibfnamefont{F.}~\bibnamefont{Sch\"{a}fer}},
  \bibnamefont{and} \bibinfo{author}{\bibfnamefont{C.}~\bibnamefont{Schmit}},
  \bibinfo{journal}{Phys. Rev. Lett.} \textbf{\bibinfo{volume}{97}},
  \bibinfo{eid}{254102} (\bibinfo{year}{2006}).

\bibitem[{\citenamefont{Huang et~al.}(2009)\citenamefont{Huang, Lai, Ferry,
  Goodnick, and Akis}}]{Huang2009}
\bibinfo{author}{\bibfnamefont{L.}~\bibnamefont{Huang}},
  \bibinfo{author}{\bibfnamefont{Y.-C.} \bibnamefont{Lai}},
  \bibinfo{author}{\bibfnamefont{D.~K.} \bibnamefont{Ferry}},
  \bibinfo{author}{\bibfnamefont{S.~M.} \bibnamefont{Goodnick}},
  \bibnamefont{and} \bibinfo{author}{\bibfnamefont{R.}~\bibnamefont{Akis}},
  \bibinfo{journal}{Phys. Rev. Lett.} \textbf{\bibinfo{volume}{103}},
  \bibinfo{pages}{054101} (\bibinfo{year}{2009}).

\bibitem[{\citenamefont{Poo et~al.}(2011)\citenamefont{Poo, Wu, Lin,
  Yang, and Chan}}]{Poo2011}
\bibinfo{author}{\bibfnamefont{Y.}~\bibnamefont{Poo}},
  \bibinfo{author}{\bibfnamefont{R.} \bibnamefont{Wu}},
  \bibinfo{author}{\bibfnamefont{Z.} \bibnamefont{Lin}},
  \bibinfo{author}{\bibfnamefont{Y.} \bibnamefont{Yang}},
  \bibnamefont{and} \bibinfo{author}{\bibfnamefont{C. T.}~\bibnamefont{Chan}},
  \bibinfo{journal}{Phys. Rev. Lett.} \textbf{\bibinfo{volume}{106}},
  \bibinfo{pages}{093903} (\bibinfo{year}{2011}).

\end{thebibliography}
\end{document}